# CHEMICAL SYSTEM COMPLEXITY AND BIFURCATION POINT: A NEW RELATIONSHIP

B. Zilbergleyt[I]

ABSTRACT

The article introduces a new relationship between the chemical system complexity and deviation of its bifurcation point from thermodynamic equilibrium. In the formalism of discrete thermodynamics of chemical equilibria, simulation of numerous equilibrium cases with regards to complexity of the reaction hosting systems has lead to a conclusion that the system deviation from thermodynamic equilibrium at the bifurcation point is directly proportional to logarithm of the system complexity parameter. With this relationship one can predict the efforts that are sufficient to destabilize the system thermodynamic branch and achieve the bifurcations area.

A relation between the system complexity and its behavior in the "far-from-equilibrium" area was qualitatively discussed in plenty of publications (e.g., [1,2,3]). Despite of some serious quantitative attempts to analyze the system complexity, perhaps the only universal conclusion with regards to chemical systems was that the more complicated is the system the more sophisticated behavior one should expect. Recently developed discrete thermodynamics of chemical equilibria [4] has introduced the Le Chatelier Response (LCR) [5,6], partly in attempt to put the system reaction to the external impact in dependence on its complexity. Being designed as Maclaurin series of the j-system shift from true thermodynamic equilibrium (TdE) $\delta_j$

(1) $$\rho_j = \Sigma_p \omega_p \delta_j^p,$$

where $p=\{0,\ldots,\pi\}$ and $\pi$ is a *chemical system complexity parameter*, the LCR has linked both values. Assuming now a proportionality between the LCR and the external thermodynamic force $F_{je}$

(2) $$\Sigma \omega_p \delta_j^p = -a_j F_{je}$$

where $a_j$ is just a coefficient, one can obtain condition of new, shifted state of chemical equilibrium at p,T=const and $\delta_j \neq 0$ as a logistic map (for full derivation see [6])

(3) $$\ln[\Pi_j(\eta_j,0)/\Pi_j(\eta_j,\delta_j)] - \tau_j \varphi(\delta_j,\pi) = 0.$$

Here the first term is a traditional expression for the Gibbs' energy change in isolated chemical system reduced by RT, $\Delta G_j/RT$. The thermodynamic equivalent of transformation $\eta_j$ is an invariant of chemical reaction, it represents amount of moles of any reaction participant, transformed in the reaction from its initial state to TdE per participant's stoichiometric unit. The numerator of the under logarithm ratio, $\Pi_j(\eta_j,0)$, corresponds to $\delta_j=0$ and equals to the reaction equilibrium constant; it is relevant to the j-system TdE and serves as a reference point for new equilibrium. $\Pi_j(\eta_j,\delta_j)$ is a regular mole fractions product of the j-reaction participants as a function of the system shift $\delta_j \neq 0$. The second term results from the system openness and is responsible for all what happens to it out of TdE. The factor $\tau_j$ is similar to the growth factor in the theory of bio-populations [7]. In general, map (3) is an expression for the system's Gibbs' free energy change that takes into account the external impact. This expression is the map of states of the chemical system because it maps TdE, the system state with $\delta_j=0$, to another state of the system with $\delta_j \neq 0$.

The weights $w_p$ are supposed to fall within the [0,1] interval; they are unknown *a priori*, and we have no other choice as to equalize all of them to unity, besides $w_0$. A new feature of this map is that the intermediate function $\varphi(\delta_j,\pi)$ and its solutions essentially depend on the $w_0$ value: they split by two groups, one for $w_0=0$ and another for $w_0 \neq 0$. Applying the method of mathematical induction for $w_0 \neq 0$ at restricted $\pi$ one can easily obtain

(4) $$\varphi_u(\delta_j,\pi)=\delta_j(1-\delta_j^{\pi}),$$

that leads to

(5) $$\ln[\Pi_j(\eta_j,0)/\Pi_j(\eta_j,\delta_j)] - \tau_j\delta_j(1-\delta_j^{\pi})=0.$$

---

[I] System Dynamics Research Foundation, Chicago, sdrf@ameritech.net

In the simplest case of $\pi=1$ map (5) turns to the basic map of state of the chemical system

(6) $$\ln[\Pi_j(\eta_j,0)/\Pi_j(\eta_j,\delta_j)] - \tau_j\delta_j(1-\delta_j)=0.$$

Applying the same to the set with $w_0=0$, we get

(7) $$\varphi_z(\delta_j,\pi)=(1-\delta_j^{\pi+1}),$$

and map (3) turns to

(8) $$\ln[\Pi_j(\eta_j,0)/\Pi_j(\eta_j,\delta_j)]-\tau_j(1-\delta_j^{\pi+1})=0.$$

Distinguished by the free term $w_0$, those system groups seem to be relevant to quite different types of chemical systems – weak, $w_0=0$, and strong, $w_0 \neq 0$; appropriate bifurcation diagrams are shown in Fig.1. They are plotted upon the simulation data for the system with reaction $PCl_3+Cl_2=PCl_5$ twice – one time in a weak and another in a strong version.

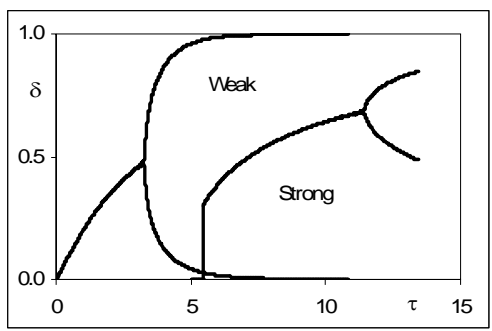

Fig.1. Weak and strong systems bifurcation diagrams, reaction $PCl_3+Cl_2=PCl_5$, $\eta=0.8$, $\pi=1$, dimensionless axes values.

The notorious "far-from-equilibrium" area starts at bifurcation point, where the system thermodynamic branch looses its stability and system becomes bi-stable. System shifts from TdE at the bifurcation points is different for the strong and for the weak systems; they reflect the efforts that should be undertaken to destabilize the systems thermodynamic branches. This critical shift value actually is one of the most important parameters defining the chemical system behavior.

To investigate a relation between the system complexity $\pi$ and the system shift from equilibrium at bifurcation point we have carried a large number of data simulations to plot bifurcation diagrams for the symbolic reactions like $aA+bB=cC$ (or briefly $\{a,b,c\}$) varying b and c from 1 to 4 in different combinations. This effort resulted in a discovery of new interesting relationship. A typical set of results for the reaction $\{1,1,-1\}$ (in this case that was $PCl_3+Cl_2=PCl_5$) is presented graphically in Fig.2.

The linearity of $\delta_{bp}$ vs. $\ln\pi$ in Fig.2 is obvious; the new empirical rule, following from the simulation results, is

(8) $$\delta_{bp} = \delta_0 + \beta \ln\pi.$$

The same results have been strongly confirmed in all simulation cases. The rule (8) states that *deviation of the chemical system from TdE, where its thermodynamic branch becomes unstable, is directly proportional to the logarithm of the system complexity coefficient.* Term $\delta_0$ corresponds to the simplest chemical system with $\pi=1$; its value not essentially depends upon $\eta$. The values of coefficients $\beta$ also slightly vary with $\eta$ but they are very close, and all lines eventually converge forming a fascia. The rule also means that *the larger is π, or the more complicated is the chemical system, the less it is prone to evolution*. Equation (8) is to replace the linguistic variable "far-from

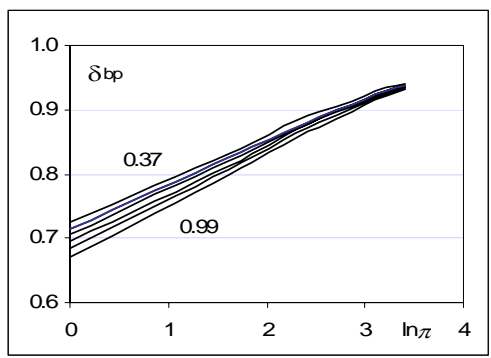

Fig.2. Shift value at the bifurcation point vs. ln$\pi$, reaction {1,1,-1}, numbers - varying η.

-equilibrium" with a precise number, using the complexity parameter; it is the locator of where that area starts in terms of the system deviation from TdE. The rule is the same for both types of chemical systems, the strong and the weak, as well as for both types of bifurcation diagrams, static, in coordinates $\delta_j$ vs. $\tau_j$ and dynamic, in coordinates $\delta_j$ vs. $F_{je}$ [6].

To the best of our knowledge, no quantitative relation between the system complexity and its resistance/proneness to bifurcations were ever described before. This newly discovered rule for chemical systems was already briefly mentioned by the author [6]; essentially more numerous research results obtained since that time served as a basis for this publication. Though we are not intended to extend it beyond the chemical systems, one can expect that many of non-chemical systems, whose evolution follows bifurcation patterns, may show the same features.

The author is grateful to S. Marshenina from the SUL Research, Ekaterinburg, Russia, for inspiring discussions.